\documentclass{PoS}

\def\mpi2{m_\pi^2}
\def\mK2{m_K^2}

\newcommand{\bea}{\begin{eqnarray}}
\newcommand{\eea}{\end{eqnarray}}
\newcommand{\be}{\begin{equation}}
\newcommand{\ee}{\end{equation}}

\def\lsim{\raise0.3ex\hbox{$<$\kern-0.75em\raise-1.1ex\hbox{$\sim$}}}
\newsavebox{\DERIVBOXZLM}
\savebox{\DERIVBOXZLM}[2.5em]{$\Longrightarrow\hspace{-1.5em}
\raisebox{.2ex}{*}
\hspace{-.7em}\raisebox{-.8ex}{\scriptsize lm}\hspace{.7em}$}

\usepackage{epsf}
\title{Status of the MILC calculation of electromagnetic contributions to pseudoscalar masses}

\ShortTitle{Electromagnetic contributions to pseudoscalar masses}

\author{S.~Basak$^a$, A.~Bazavov$^b$, C.~Bernard$^c$, C.~DeTar$^d$, E.~Freeland$^e$, W.~Freeman$^f$,
 J.~Foley$^d$, Steven~Gottlieb$^g$, U.M.~Heller$^h$, J.E.~Hetrick$^i$, J.~Laiho$^j$, \speaker{L.~Levkova}$^{,d}$,
 M.~Oktay$^d$, J.~Osborn$^k$, R.L.~Sugar$^l$, A.~Torok$^g$, D.~Toussaint$^m$, R.S.~Van~de~Water$^b$, R.~Zhou$^g$ \,(MILC Collaboration)\\

$^a$ NISER, Bhubaneswar, Orissa 751005, India      \\
$^b$ Physics Department, Brookhaven National Laboratory, Upton, NY 11973, USA\\
$^c$  Physics Department, Washington University,
St. Louis, MO 63130, USA\\
$^d$ Physics Department, University of Utah,
Salt Lake City, UT 84112, USA\\
$^e$ Department of Physics, Benedictine University, Lisle, IL 60532, USA\\
$^f$ Department of Physics, George Washington University, Washington, DC 20037, USA\\
$^g$ Department of Physics, Indiana University,
Bloomington, IN 47405, USA\\
$^h$ American Physical Society, One Research Road, Ridge, NY 11961, USA\\
$^i$ Physics Department, University of the Pacific,
Stockton, CA 95211, USA\\
$^j$ SUPA, School of Physics and Astronomy, University of Glasgow, Glasgow, UK\\
$^k$ ALCF, Argonne National Laboratory,
Argonne, IL 60439, USA\\
$^l$ Physics Department, University of California,
Santa Barbara, CA 93106, USA\\
$^m$ Physics Department, University of Arizona
Tucson, AZ 85721, USA\\
E-mail: \email{ludmila@physics.utah.edu}
}
\abstract{ We calculate pseudoscalar masses on gauge configurations containing
the effects of  2+1 flavors of dynamical asqtad quarks and quenched
electromagnetism.  The lattice spacings vary from 0.12 to 0.06 fm. 
The masses are fit with staggered chiral perturbation theory including NLO electromagnetic terms.
We attempt to extract the fit parameters for the electromagnetic contributions,
while taking into account the finite volume effects, and extrapolate them to the physical limit.}

\FullConference{The XXX International Symposium on Lattice Field Theory\\
                 July 24-29, 2012\\
                 Cairns, Australia}

\begin{document}

\section{Introduction}
Quark masses are fundamental parameters in the Standard Model and their
determination from lattice field
theory is important for phenomenology.
With the advance of high-precision spectroscopy the electromagnetic (EM) contributions
to the quark masses have become important. In fact, they constitute the largest uncertainty
in the calculation of $m_u/m_d$ \cite{qrat} as indicated in Table~\ref{tab:masses}.
\begin{table}[b]
\begin{center}
\begin{small}
\begin{tabular}{lccc}\hline\hline
& $m_u$ [MeV] & $m_d$ [MeV] & $m_u/m_d$\\
\hline\hline
value & 1.9&4.6 & 0.42\\
statistics& 0.0 &0.0 &0.00\\
lattice systematics & 0.1 &0.2 &0.01\\
perturbative &0.1 &0.2& --\\
EM& 0.1& 0.1&0.04\\
\hline\hline
\end{tabular}
\end{small}
\label{tab:masses}
\caption{Value of $m_u/m_d$ calculated by the MILC Collaboration \cite{qrat} and  
estimates of the different sources of error (with $m_u$ and $m_d$ determined in the $\overline{MS}$ scheme at 2 GeV).
The row labeled ``EM'' shows the size of the
error coming from a phenomenological
estimate of electromagnetic contributions to the quark~masses.}
\end{center}
\end{table}
The EM error stems from uncertainties in the size of
the EM contributions 
to the masses of the $\pi$ and especially the $K$ mesons.
The contributions have until recently been taken from a variety of
phenomenological
estimates, and therefore have quite large and not well controlled errors.
The MILC collaboration has been working on reducing these uncertainties and
progress has been reported previously in Refs.~\cite{basak,schpt,torok}.

To lowest order (LO) in chiral perturbation theory ($\chi PT$) the electromagnetic splitting in the pion and kaon systems
is identical (this observation is known as Dashen's theorem \cite{dashen}).
We aim to calculate on the lattice the corrections to Dashen's theorem 
by employing staggered $\chi PT$ 
 with EM effects for fits to spectrum data.
The corrections to Dashen's theorem may be parametrized as~\cite{Colangelo:2010et}:
\be
 (M^2_{K^\pm}-M^2_{K^0})^\gamma =(1+\epsilon) (M^2_{\pi^\pm}-M^2_{\pi^0})^\gamma\ ,
\ee
where the superscript $\gamma$ denotes the EM contributions.\footnote{We do 
not include quark-mass renormalization effects due to the presence of the EM field, since they are
much smaller than our precision.}
Our main motivation for computing  $\epsilon$
is to be able to remove the EM effects from the experimental $K$ masses; the resulting
pure-QCD masses may be compared with QCD simulations in order to 
extract $m_u/m_d$.
Staggered $\chi PT$, developed in Ref.~\cite{schpt}, gives for the EM 
splitting between Goldstone pseudoscalars with charged and neutral quarks,
$\Delta M^2_{xy,5}$, the following NLO expression:
\bea
\Delta M^2_{xy,5}&=&q^2_{xy}\delta_{EM}-\frac{1}{16\pi^2}e^2q^2_{xy}M^2_{xy,5}\left
[3\ln(M^2_{xy,5}/\Lambda^2_\chi)-4\right]\nonumber\\
&&-\frac{2\delta_{EM}}{16\pi^2f^2}\frac{1}{16}\sum_{\sigma,\xi}\left[q_{x\sigma}q_{xy}l(M^2_{x\sigma,\xi})
-q_{y\sigma}q_{xy}l(M^2_{y\sigma,\xi})\right]
 + ({\rm {\footnotesize analytic\; terms}})
\label{eq:fit}
\eea
where $\sigma$ is the sea-quark label, $\xi$ is the staggered taste, $q_{xy}$ is the charge of a meson made 
of a quark $x$ and an antiquark $y$ (for more details on the notations see 
Ref.~\cite{schpt}).
To fit our spectrum data, we use the above expression plus the NNLO analytic terms. 
For investigating the finite volume effects, we include standard terms
dependent on $m_\pi L_s$ from EM tadpoles, as well as 
an empirical EM finite volume correction in the form $f_vq_{xy}^2/L_s^2$ used previously in Ref.~\cite{BMW},  
where $f_v$ is a constant and $L_s$ is the spatial lattice size.
However, as discussed below, our measured finite volume effects are rather small at present, and
including or omitting the finite volume terms from the fits currently makes little difference in
the final results.

\section{Lattice setup}
Our EM field on the lattice is ``photon-quenched,'' {\it i.e.},
the sea quarks are taken to be electromagnetically neutral, and 
the $U(1)_{\rm EM}$ links are generated independently of the $SU(3)_{\rm QCD}$ links. The fact that
the photons are not present in the sea does not affect the NLO fits to the
EM meson splittings since, as shown in Ref.~\cite{BD}, the unknown LEC controlling
sea-quark mass dependence cancels in EM splittings between mesons with the same
valence masses. There are physical NLO effects missing due to the absence of sea-quark charges,
but these may be inserted after the fact since they appear only in chiral logarithm terms with
known (LO) coefficients.
Uncontrolled EM quenching effects do come in at NNLO, but should be negligible at the current
level of precision.
Because diagrams that have intermediate gluon or photon states (``quark-disconnected'' diagrams)
are difficult to compute on the lattice, and because of
the photon quenching, we have to redefine the neutral pion on the lattice $^\prime\pi^{0\prime}$ 
to have the mass:
\be
m^2_{^\prime\pi^{0\prime}}=(m^2_{u\bar u} + m^2_{d\bar d})/2,
\ee
where $u\bar u$ and $d\bar d$ are states with charges ($2/3e$, $-2/3e$)  and ($1/3e$, $-1/3e$)
and disconnected diagrams are dropped by fiat.
The effect of the disconnected diagrams is expected to be small, since EM contributions
to the neutral pion must in any case vanish in the chiral limit.

We calculate the meson spectrum on the set of asqtad ensembles with 2+1 flavors listed in Table~\ref{tab:ensembles} \cite{qrat}.
\begin{table}[t]
\begin{center}
\begin{small}
\begin{tabular}{ccclr}
\hline\hline
$\approx a$[fm]& Volume & $\beta$ & $m_l/m_s$& \# configs.\\
\hline\hline
0.12 & $20^3\times64$ & 6.76& 0.01/0.05& 2254\\
      & $28^3\times64$ & 6.76& 0.01/0.05& 274\\
      & $20^3\times64$ & 6.76& 0.007/0.05& 1261\\
      & $24^3\times64$ & 6.76& 0.005/0.05& 2099\\
\hline\hline
0.09  & $28^3\times96$ & 7.09& 0.0062/0.031& 1930\\
      & $40^3\times96$ & 7.08& 0.0031/0.031& 1015\\
\hline\hline
0.06  & $48^3\times144$ & 7.47& 0.0036/0.018& 670\\
\hline\hline
\end{tabular}
\end{small}
\end{center}
\caption{Parameters of the asqtad ensembles used in this study.}
\label{tab:ensembles}
\end{table} 
The values of the electron charge that we use are $e= \pm0.909,\pm0.606$ and,$\pm0.303,0$, corresponding to
$9\alpha_{phys}$, $4\alpha_{phys}$, $\alpha_{phys}$ and 0. 

\section{Results}
\begin{figure}[b]
\epsfxsize=0.65\textwidth
\begin{center}
\epsfbox{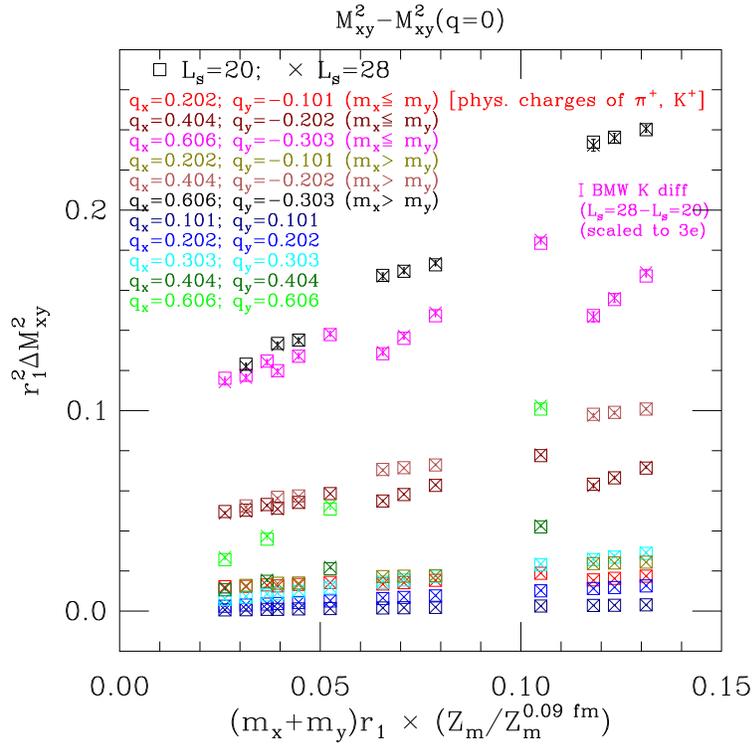}
\end{center}
\vspace{-0.5cm}
\caption{EM pseudoscalar splittings calculated on two coarse ensembles with $m_l/m_s=0.01/0.05$
 and spatial sizes $L_s=20$ and $L_s=28$ {\it vs.} the sum of the valence quark masses. The results for the different
quark charge combinations are distinguished by color.   
The size of the finite volume effects found
by the BMW group \cite{BMW} is shown in magenta in the upper right and is scaled to charge $3e$, to correspond
to our magenta points plotted slightly below that. The finite volume effects we see (difference between
crosses and squares) are smaller than those found by BMW.}
\label{fig:FV}
\end{figure}

\begin{figure}[t]
\epsfxsize=0.7\textwidth
\begin{center}
\epsfbox{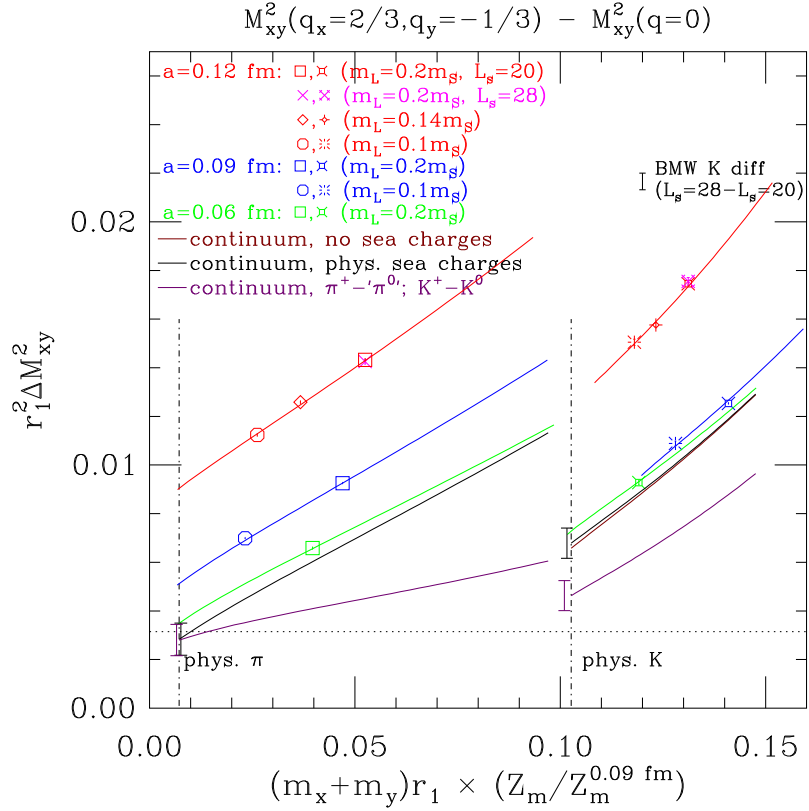}
\end{center}
\vspace{-0.5cm}
\caption{Our preliminary results for electromagnetic corrections in pseudoscalar mesons. Plotted is the
difference in the squared mass of mesons with charged valence quarks and neutral valence quarks {\it vs.} the
sum of the valence quark masses. Only a small subset of the data that was used in the fit is shown in the plot.
This fit had 55 data points, 26 parameters, and
an (uncorrelated) $p$ value of 0.09.
The red, blue and green curves correspond to three different lattice spacings and include the effects of taste
breaking. The vertical dot-dashed lines correspond to the quark masses of the $\pi$ and $K$ mesons. The purple
curves are the continuum limits for $K^+-K^0$,
and for $\pi^+-^\prime\pi^{0\prime}$
(left), where the $^\prime\pi^{0\prime}$ is the average
of a neutral pion with charges ($2/3e$,$-2/3e$) and ($1/3e$,$-1/3e$). The latter is an approximation to the physical $\pi^0$,
which would also be affected by quark annihilation terms not included in our calculation. The horizontal
dotted line is the physical value of the pion electromagnetic mass splitting and is in close agreement with our
calculation. The result for the $K$ is larger than for the pion, giving a correction to the lowest order of Dashen's
theorem by a factor of $\epsilon=0.65(7)$.  In the upper right we show the expected size of finite volume
effects for the kaon from the BMW calculation \cite{BMW}.  It should be compared with the (tiny) difference between
the red square and magenta cross, or 'fancy' red square and 'fancy' magenta cross.}
\label{fig:EMfit}
\end{figure}

We do not see large finite volume effects in the EM splittings calculated on the $a\approx0.12$~fm 
ensembles with $L_s=20$ and 28. 
The volume effects do increase with the quark charges, but remain small.
In Fig.~\ref{fig:FV} we compare our results with the finite volume effects 
for a charged kaon  (with 3 times the physical charges) expected from the 
BMW collaboration result \cite{BMW} (magenta error bar).
We find that ours are about three times smaller. 
A similar comparison is shown in Fig.~\ref{fig:EMfit} for physically charged
kaons.  Nevertheless, the error on the finite volume effect in our data (computed simply by propagating the errors in the difference between the results on the two volumes) is slightly larger than 
the effect itself,
so any tension with the BMW result is less than a $2\sigma$ effect. Indeed, we see
no reason 
for any difference with BMW other than statistics,
since as far as we can tell, we are treating the $U(1)_{\rm EM}$ fields in the same way they do,
and we do not expect the EM finite volume effects to be strongly dependent on quark mass or quark action.
The finite volume effects we see are
also smaller than those expected from the RBC Collaboration results \cite{Blum:2010ym}, 
but in that case some or all  of the difference may be
due to the different treatment of zero modes in our $U(1)_{\rm EM}$ fields.
We are currently increasing the statistics on the  $a\approx0.12$ fm, $L_s=28$ ensemble in hopes
of clarifying this issue.

\begin{figure}[t]
\epsfxsize=0.7\textwidth
\begin{center}
\epsfbox{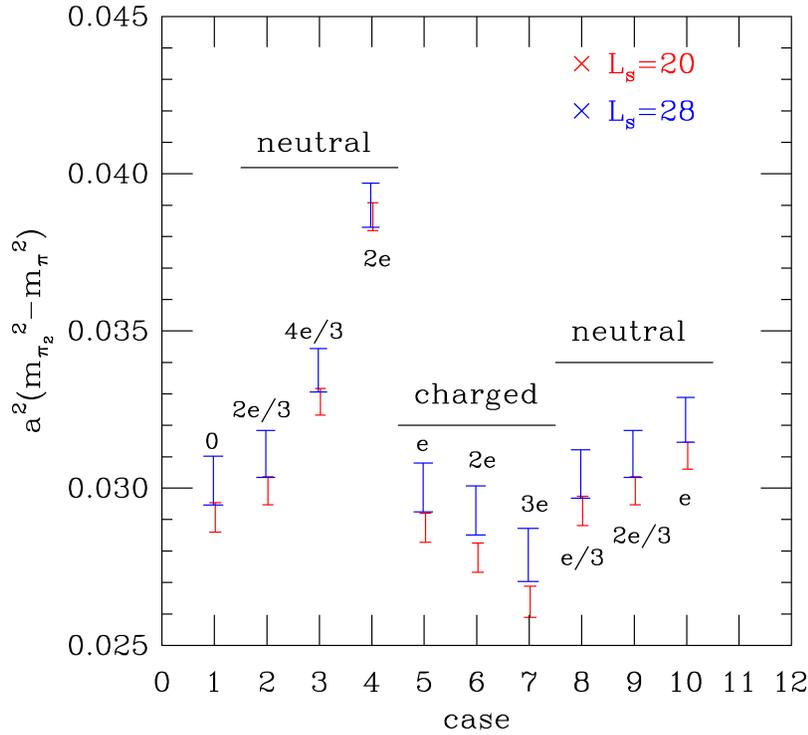}
\end{center}
\vspace{-0.5cm}
\caption{ Taste splitting between the Goldstone pion and the second local pion on the coarse ensembles
with $m_l/m_s=0.01/0.05$ and $L_s=20$ and $L_s=28$.  The $x$-axis values are
arbitrary labels
for the mesons with various values of the  quark charges. 
Case 1 is a pion made from uncharged quarks.
The first group of neutral pions (cases 2--4) 
are essentially
$\bar u u$ combinations, and the second neutral group (cases 8--10) are of the $\bar d d$ type (the ($2e/3$, $-2e/3$) combination appears 
in both neutral groups). These groups are labeled by one of the quark charges (the other has an opposite sign).  
The middle group (cases 5--7), has quark charges with the physical ratio of (2:1); they are labeled by the
total meson charge.
}
\label{fig:taste}
\end{figure}

The taste splitting between the Goldstone pion and the second local pion
grows with the quark charge for the neutral pions (see Fig.~\ref{fig:taste} for the two
coarse ensembles with the same parameters and different $L_s$ in Table~\ref{tab:ensembles}). 
This shows that, for charges larger than the physical charges, quark taste changing by 
high-momentum photons becomes a significant correction to that by high-momentum gluons. 
For charged pions, where annihilation into photons is forbidden, the effect is smaller
(and negative), but still evident. 
 The fit function Eq.~(\ref{eq:fit}) is based
on the neglect of taste-violations caused by photons. For that reason,
the analysis presented here  focuses only on
fits with physical quark charges.  However, to the extent that photon-induced
taste violations remain small compared with gluon-induced violations, which is
arguably the case for all values of quark charges shown in Fig.~\ref{fig:taste}, one should be
able to expand the fit function in powers of $\alpha_{EM}$. This would allow
such effects to be included by adding $\alpha^2_{EM}$ analytic terms to the fit function.  
We have made some preliminary fits along these lines to our full data set with quark
charges up to three times the physical values, and they seem to work reasonably well.

Before leaving the subject of taste violations, we note that
Fig.~\ref{fig:taste} shows that taste splitting 
is not very much affected by finite volume effects
(compare the data in blue and red at $L_s=20$ and 28 for the above coarse ensembles).

We now fit, to Eq.~(\ref{eq:fit}) plus NNLO analytic terms, the EM splitting of the meson spectrum using the data with the physical
electron charge $e$ (i.e., mesons made from combinations of $u$ and/or $d$ quarks,
with $q_u=2/3e$ and $q_d=-1/3e$). Partially quenched points are included; in typical
fits there are between 50 to 120 data points, with 20 to 30 fit parameters (depending
on how many of the NNLO terms are included, and whether small variations with $a^2$
of the LO and NLO low energy constants are allowed). A typical fit is shown
in Fig.~\ref{fig:EMfit}, along with the unitary pions and kaons on each ensemble. 
The fit is uncorrelated:
the covariance matrix is nearly singular, and the
statistics are insufficient to determine it with enough precision to
yield good correlated fits. Based on the fits, our preliminary determination of 
the correction
to Dashen's theorem in the continuum limit and at physical quark mass is
$\epsilon=0.65(7)(14)$,
where the errors are statistical and systematic, respectively.  The systematic
error comes largely from variations in $\epsilon$ when the assumptions going
into the chiral/continuum fit are changed.
Systematic
 effects are still under investigation.  In particular, based on the currently small
finite volume effects seen in our data, we do not include a finite volume error here.  However,
we suspect that better statistics on our larger lattices will yield a statistically significant finite
volume error.
Our result is compatible with previous ones: 0.60(14) (statistics only) \cite{Portelli:2010yn},
0.628(59) (statistics only) \cite{Blum:2010ym}, and
0.70(4)(8) \cite{BMW}.
With our value for $\epsilon$ from above, our preliminary estimate for the EM uncertainty in $m_u/m_d$
is reduced to 0.022 \cite{doug}, which is about half of our previous error shown in Table~\ref{tab:masses}.

{\bf Acknowledgements:} Computations for this work were carried out with resources provided by the USQCD
Collaboration, the Argonne Leadership Computing Facility, and the National Energy Research Scientific 
Computing Center, which are funded by the Office of Science of the U.S.
Department of Energy; and with resources provided by the National Center for Atmospheric
Research, the National Institute for Computational Science, the Pittsburgh Supercomputer
Center, the San Diego Supercomputer Center, and the Texas Advanced Computer Center,
which are funded through the National Science Foundation's XSEDE Program. This work
was supported in part by the U.S. Department of Energy under Grants DE-FG02-91ER-
40628, DE-FG02-91ER-40661, DE-FG02-04ER-41298, and DE-FC02-06ER41446; and by
the National Science Foundation under Grants PHY07-57333, PHY07-03296, PHY07-57035,
PHY07-04171, PHY09-03571, PHY09-70137, and PHY10-67881. This manuscript has been
co-authored by an employee of Brookhaven Science Associates, LLC, under Contract No.
DE-AC02-98CH10886 with the U.S. Department of Energy. Fermilab is operated by Fermi
Research Alliance, LLC, under Contract No. DE-AC02-07CH11359 with the U.S. Department of Energy. 
For this work we employ QUDA \cite{quda}.   

\end{document}